# Investigation on the Compressibility Characteristics of Low Mach Number Laminar Flow in Rotating Channel

Junxin Che[1,2,3], Ruquan You[1,2,3,*], Wenbin Chen[4], Haiwang Li[1,2,3]

1 National Key Laboratory of Science and Technology on Aero Engines Aero-thermodynamics, Beihang University, Beijing 100191, China

2 Research Institute of Aero-Engine, Beihang University, Beijing 100191, China

3 Frontiers Science Center for Super-cycle Aeroengine's Aerothermodynamics, Beihang University, Beijing 100191, China

4 Hunan Key Laboratory of Turbomachinery on Small and Medium Aero-Engine, AECC Hunan Aviation Powerplant Research Institute, Zhuzhou 412002, China

## Abstract

In high-speed rotating channels, significant compressive effects are observed, resulting in distinct flow characteristics compared to incompressible flows. This study employs a finite volume method, based on the implicit formulation, to solve for low-speed compressible laminar flow in rotating channels using an orthogonal uniform grid. The governing equations include the full Navier-Stokes equations and the energy equation. Contrary to stationary channel, the alterations in flow within rotating channel are primarily influenced by the compressive effects of centrifugal force and the compressibility of fluid within the flow's normal section. The first effect involves a reduction in the velocity due to centrifugal force, leading to an increasing influence of the Coriolis force compared to inertial forces. This trend changes aligns closely with the increase in rotation speed. The second effect arises from the increase in Mach number and the Coriolis compression, resulting in slight density differences within the cross-section. Strong centrifugal forces generate significant centrifugal additional force (buoyancy force). Consequently, under the same local rotation number, the velocity profiles of the mainstream experience considerable changes. Additionally, higher Mach number significantly impact wall shear stress, with the leading side being notably affected. For instance, at a cross-sectional $Ro = 0.6$ and $Ma = 0.035$, the dimensionless shear stress on the leading side decreased by 13%. Furthermore, while an increase in Mach number has minimal impact on the cross-sectional secondary flow structure, changes in mainstream velocity profiles influence secondary flow intensity, resulting in an enhanced velocity peak and a shift towards the trailing side.

## Nomenclature

| | | |
|---|---|---|
| | $D$ | Hydraulic diameter (m) |
| | $r$ | Rotational radius (m) |
| | $x$ | Spanwise direction length (m) |
| | $y$ | Normal direction length (m) |
| | $z$ | Streamwise direction length (m) |
| | $\mu$ | Molecular dynamic viscosity [kg/(m s)] |
| | $\rho$ | Density (kg/m³) |
| | $p$ | Pressure (Pa) |
| | $\Omega$ | Speed of rotation (rad/s) |
| | $u$ | velocity (x axis) (m/s) |

* Corresponding author at: National Key Laboratory of Science and Technology on Aero Engines Aero-thermodynamics, Beihang University, Beijing 100191, China.
E-mail address: youruquan10353@buaa.edu.cn (R. You).

| | | |
|---|---|---|
| $v$ | velocity (y axis) (m/s) | |
| $w$ | velocity (z axis) (m/s) | |
| $e$ | Internal energy | |
| $k$ | Heat capacity ratio of air (-) | |
| $\nabla$ | Hamiltonian | |
| $\boldsymbol{u}$ | vector of velocity (m/s) | |
| $\boldsymbol{e_y}$ | unit vector of y axis (-) | |
| $Re$ | Reynolds number ($\rho u D/\mu$) (-) | |
| $Ro$ | rotation number ($\Omega D/u$) (-) | |
| $Ma$ | Mach number ($u/\sqrt{kRT}$) (-) | |
| $CW$ | Centrifugal Work number ($Ro^2 Ma^2 r/D$) (-) | |
| $Buo$ | Buoyancy number (($r/D$)$Ro^2$($\Delta\rho/\rho$)) (-) | |
| $f$ | dimensionless wall shear stress (-) | |
| $\tau_w$ | wall shear stress (Pa) | |
| $LS$ | Leading side | |
| $TS$ | Trailing side | |

# 1 Introduction

In the 1950s, the emergence of rotating machinery spurred profound investigations into flow dynamics within rotating channels. Unlike stationary conditions, flow within these channels is influenced by an interplay of various rotational forces, encompassing centrifugal, Coriolis, and buoyancy forces. These forces collaboratively act upon fluid elements, accentuating the intricacies of flow within rotating channels.

Laminar flow within rotating channels exhibits simpler influencing factors compared to turbulent flow, leading to earlier initiation of mechanistic research due to its lower complexity. Generally, fluid Mach number within rotating channels remain notably below 0.3, indicating incompressible flow. Therefore, research on rotating channel flow operates within the framework of incompressibility. Systematic research on incompressible laminar flow within rotating channels was conducted as early as the 1970s. These research primarily focused on the development of the main velocity profile and secondary flow structures. Furthermore, the stability of laminar flow has received comprehensive attention.

Johnston (1972)[1] and Kristoffersen (1993)[2] examined the influence of the Coriolis force on the velocity profiles of rotating channel cross-sections through experiments and direct numerical simulations. The Coriolis effect on the mean flow velocity profile in a rotating channel disrupts the symmetrical distribution, causing a shift in the maximum velocity towards the high-pressure side. Alfredsson and Persson (1989)[3] investigated flow within rotating channels at higher Reynolds number and found that the roll-cell structure initially emerges in the downstream region. With an increasing rotation rate, the roll-cell structure progressively shifts upstream. Larger rotation number can trigger the transition of this roll-cell structure into turbulence. Experimentally, it was observed that turbulence could be induced at Reynolds number as low as 600.

Laminar secondary flow within rotating channels flow displays distinctive features. When stationary, Poiseuille flow exhibits pronounced symmetry. At low rotation number, fully developed laminar flow presents a pair of vortex structures in its cross-sectional secondary flow. Lezius (1976)[4] and collaborators noted that as the rotation rate increases, the stability of laminar flow deteriorates, leading to an unstable regime. However, as the rotation rate continues to rise, the flow reverts to a stable state. It is generally understood that the stable state's critical points are around $Re$ = 88 and Ro = 0.5. This signifies that when the Reynolds number is lower than this threshold, the flow *Re*mains consistently stable. This phenomenon has been observed in channels of various aspect ratios. Along the channel's axial centerline, velocity often exhibits a wavering pattern under unstable conditions.

In the case of higher rotational speeds in rotating channels, Speziccle (1982)[5] observed the decomposition of the double-vortex secondary flow into four asymmetric counter-rotating vortex configurations, similar to those observed in highly curved rectangular channels with elevated Dean number. Additionally, it was initially demonstrated that with further increases in rotation rate, the flow would re-stabilize into a slightly asymmetrical double-vortex configuration, exhibiting a Taylor–Proudman regime within the channel. In a channel with an aspect ratio of 8, Speziccle (1983)[6] investigated the secondary flow characteristics resembling nearly rectangular channel flow. The cross-sectional secondary flow originates from the corner regions and gradually progresses toward the channel's center. As the rotation rate increases, the large-scale double-vortex secondary flow within the channel gradually bifurcates into multiple counter-rotating vortex structures.

The aforementioned studies were conducted under non-heated wall conditions. In incompressible flow, the fluid density within the channel remains essentially constant, and the rotational volume force along the flow direction can be equivalently accounted for as a static effect by introducing a constant gradient to the pressure field. However, when the walls are heated, a significant density disparity emerges across the cross-section of the rotating channel. As a consequence of this density variation, the centrifugal force component, known as buoyancy force, exerts a significant influence on the flow dynamics.

Mori (1971)[7] theorized that in fully developed flow within a rotating channel, the Coriolis-induced secondary flow significantly enhances flow resistance and heat transfer capacity. In turbulent flows, the increase in resistance and heat transfer is less pronounced than in laminar flows. Morris (1979)[8] discovered that for radially outward flow, the Coriolis acceleration often enhances heat transfer. However, this enhancement might be offset and reversed by the centrifugal buoyancy effect. Utilizing numerical techniques, Siegel (1985)[9] examined the impact of buoyancy on heat transfer in a rotating laminar channel. For radially inward flow, buoyancy often improves heat transfer, while for radially outward flow, it tends to diminish it. Song (1990)[10] found that centrifugal buoyancy substantially reduces wall heat transfer in outward-flowing rotating channels. Conversely, in inward-flowing rotating channels, this effect enhances heat transfer. Fann (1992)[11] conducted a numerical study on the velocity and temperature fields within a heated laminar channel. Rotation led to a significant increase in the Nusselt number on the downstream wall and a slight increase on the side walls, with a slight decrease on the upstream wall. This resulted in temperature decreases on the downstream wall and increases on the upstream wall. Generally, with an increasing rotation number, both the average flow resistance coefficient and Nusselt number of the channel increase. Yan (1995)[12] and colleagues investigated heat transfer and flow resistance characteristics in rotating channels with both constant wall temperature and constant heat flux boundary conditions. They found that near the entrance, the Nusselt number increased with an increasing Reynolds number, but the trend reversed as the fluid progressed downstream. The average flow resistance coefficient and Nusselt number for constant wall temperature conditions were lower than those for constant heat flux conditions near the entrance, but this trend reversed farther downstream. Han (1992)[13] also studied variations in heat transfer characteristics with different thermal boundary conditions using the copper block heating method. Under a given buoyancy parameter, the Nusselt number on the downstream surface under constant heat flux conditions was 10-20% higher than under constant wall temperature conditions, while the Nusselt number on the upstream surface under constant heat flux conditions was 40-80% higher than under constant wall temperature conditions for the same buoyancy parameter. Hwang (1998)[14] and colleagues discovered that the Coriolis force induced by rotation significantly increases wall friction losses in straight channels. However, the head losses in sharp bends decrease with an increasing rotation number.

In recent years, research on turbulence in rotating channels has become increasingly deep. Regarding the impact of rotation on turbulence, Nakabayashi(2005)[15] and Brethouwer(2016)[16](2017)[17] found that even in high Reynolds number channels, the leading side begins to transition to laminar flow as the rotation number increases. However, at high Reynolds number and high rotation speeds, linearly unstable Tollmien-Schlichting-like waves can cause sustained bursting of turbulence. Grundestam(2008)[18] discovered that a balance between negative turbulence production mechanisms played a significant role in maintaining the relatively distinct interface between the two regions. Khan(2021)[19] developed a nonlinear theory to describe the laminar to turbulent transition of a rotating Poiseuille flow under the influence of the Coriolis force. Xia (2016)[20](2018)[21], Yang (2021)[22] and Zhang (2022)[23] conducted a comprehensive systematic study on rotating channel Turbulence, which includes investigations of mean velocity characteristics, the multiple states in turbulence and the flow structures in rotating plane Poiseuille flow(plumes, inertial waves, and plume currents).

With the development of flow visualization technology, there is a clearer understanding of the flow characteristics in the rotating channel. Liou(2003)[24] used LDV to obtain spectral analysis of flow in rotating channels. Sante(2010)[25] and Visscher(2011)[26] observed an increase in boundary layer thickness near the leading side and a decrease in boundary layer velocity near the trailing side under rotational conditions using PIV. Coletti (2014)[27] studied the flow characteristics in a rotating ribbed channel by time-resolved particle image velocimetry. When the rib wall is heated, the near-wall layer is subjected to significant centripetal buoyancy. Due to the combined action of the rotational Coriolis force and the buoyancy force, the recirculation region behind the ribs extends over the entire intercostal distance, and even extends across the entire intercostal space. You (2018)[28] measured the velocity field in the rotating heating channel by PIV and captured the flow separation phenomenon near the leading side wall caused by buoyancy.

In a previous study by Che (2023)[29], the flow within cooling channels of rotating machinery under realistic operating conditions was investigated. The results indicated that under high-speed rotation, the centrifugal force-induced compression effect significantly alters the flow within the channels. Through theoretical derivation and dimensional analysis, we identified a new significant parameter - the centrifugal work number ($CW$). This suggests that the flow within rotating channels in real rotating machinery cannot overlook the influence of fluid compression effects.

The current research on flow within rotating channels primarily focuses on the impact of the Coriolis force and buoyancy on flow and heat transfer mechanisms, within the framework of incompressible flow. The influence of buoyancy is also extensively studied under wall heating conditions. However, in various rotating systems, the intense centrifugal potential field generated by high rotational speeds can lead to noticeable compressibility effects within the fluid, even at relatively low Mach number. Building on existing theoretical foundations, this study employs numerical methods to investigate laminar flow within rotating channels, further delving into the mechanisms behind significant changes in the velocity profiles due to compressible flow effects at low Mach number.

## 2. Governing equations and numerical description

### 2.1 Equations for Rotating Channel Flow

The fluid governing equations for the fluid within a rotating channel can be formulated as follows:

$$\left. \begin{aligned} &\nabla \rho \boldsymbol{u} = 0 \\ &\rho(\boldsymbol{u} \cdot \nabla)\boldsymbol{u} = -\nabla p + \mu \nabla^2 \boldsymbol{u} - \rho[2\boldsymbol{\Omega} \times \boldsymbol{u} + \boldsymbol{\Omega} \times (\boldsymbol{\Omega} \times \boldsymbol{r})] \\ &\nabla \left[ \left( \rho e + \rho \frac{u^2}{2} + p \right) \boldsymbol{u} \right] = \nabla(\lambda \nabla T) - \rho \boldsymbol{\Omega} \times (\boldsymbol{\Omega} \times \boldsymbol{r}) \cdot \boldsymbol{u} \end{aligned} \right\} \quad (2.1)$$

An essential application of similarity theory in fluid mechanics is guiding experimental setups and organizing experimental data. Within the framework of similarity theory, results obtained from individual experiments can represent the entire flow within models sharing the same dimensionless parameters. Additionally, in investigating specific flow and heat transfer phenomena, experimental arrangements should be guided by similarity criteria. By non-dimensionalizing the aforementioned governing equations, the dimensionless equations can be expressed as follows:

$$\left. \begin{aligned} &\nabla \rho' \boldsymbol{u}' = 0 \\ &\rho'(\boldsymbol{u}' \cdot \nabla')\boldsymbol{u}' = -\nabla' p' + \frac{1}{Re} \nabla'^2 \boldsymbol{u}' + Ro\left(2\boldsymbol{e}_y \times \boldsymbol{u}'\right) - Ro^2 \frac{r_0}{D_0} \left(\boldsymbol{e}_y \times \left(\boldsymbol{e}_y \times \boldsymbol{r}'\right)\right) \\ &\nabla'\left(\rho' c_p' \theta' \boldsymbol{u}'\right) + (k-1) Ma^2 \nabla'\left(\rho' \frac{u'^2}{2} \boldsymbol{u}'\right) = \frac{1}{RePr} \nabla'(\nabla' \theta') - (k-1) CW \rho'\left(\boldsymbol{e}_y \times \left(\boldsymbol{e}_y \times \boldsymbol{r}'\right)\right) \boldsymbol{u}' \end{aligned} \right\} \quad (2.2)$$

In cases where the walls are unheated and the density of the fluid across the cross-section can be approximated as constant, the pressure gradient can effectively counterbalance the influence of centrifugal forces. Here, the centrifugal force term can be combined with the pressure term, leading to the fusion of static pressure and the centrifugal force field's function, resulting in a combined pressure term:

$$p^* = p - \rho_0 \frac{\Omega^2 r^2}{2} \tag{2.3}$$

In this case, the key parameter, $Ro^2 \frac{r_0}{D_0}$, derived from the centrifugal force term in the momentum equation loses its significance. Simultaneously, the work done by the centrifugal force term in the energy equation can be disregarded.

However, when the compressibility effect of the rotational centrifugal force on the fluid within the channel becomes significant, another crucial dimensionless parameter, denoted as $CW$, $CW = Ro^2 \frac{r_0}{D_0} Ma^2$, can be deduced from the work done by the centrifugal force term in the energy equation. The specific derivation of this parameter and its effects can be referenced from Che (2023)[29]. $CW$ represents the ratio of the work done by the centrifugal force to the gas enthalpy value, effectively characterizing the compressibility effect of the centrifugal force on the fluid. Generally, when the $CW$ is small, the compressibility effect of the centrifugal force on the fluid can be practically ignored, allowing the fluid to be treated as incompressible. However, when the $CW$ number is large, the work done by the centrifugal force on the fluid becomes substantial. In this cases, as the fluid experiences a rise in pressure, mechanical energy is transformed into internal energy, leading to an increase in fluid temperature along its path.

## 2.2 Numerical set-up

To delve into the flow characteristics within a rotating channel under the notable influence of centrifugal compression, our study employs a small-scale model featuring a hydraulic diameter of 1 mm, as depicted in Figure 1. This selection enables us to attain higher flow velocity within the channel, maximizing the Mach number under laminar flow conditions. The four walls of the channel are designated as adiabatic.

By maintaining consistent crucial parameters such as Reynolds number and rotation number, we investigate the impact of varying channel inlet Mach number by altering the inlet fluid density. This approach enables us to explore the compressibility effects of the fluid within a rotating channel where the key parameters remain uniform.

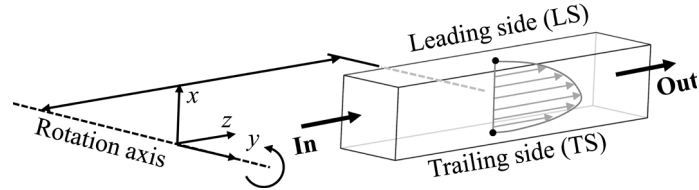

Figure 1: Schematic of the Research Model

In numerical simulations, a Coupled Solver is employed to solve the flow within the rotating channel. This solver directly tackles the transient Navier-Stokes equations, transforming the steady-state problem into a time-advancing transient problem, progressively reaching a converged steady-state solution. The equation is solved utilizing the second-order upwind scheme.

To validate the reliability of the computational approach, the velocity profiles are compared with the findings of Speziale (1982). The model exhibits a Reynolds number of 85, a rotation number of 0.27, and an aspect ratio of 2. As depicted in Figure 2, the velocity profiles obtained through the employed laminar flow solution method closely align with the literature results, affirming the credibility of our research methodology.

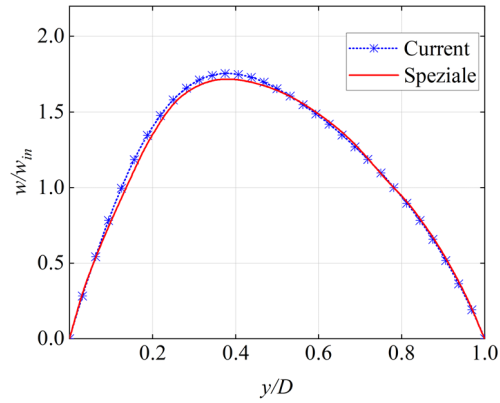

Figure 2: Numerical Method Validation

### 2.3 Validation of Grid Independence

Given the laminar nature of the flow, an equidistant grid was employed for numerical simulation. The grid refinement ranged from 10×10×100 to 50×50×500 across the cross-section, yielding the mid-section velocity profiles displayed in Figure 3. As observed, with grid refinement, the main flow velocity profiles progressively converge. The velocity discrepancy between the 30×30×300 and 50×50×500 grid configurations is less than 1%, suggesting that the 30×30×300 grid can be considered grid-independent.

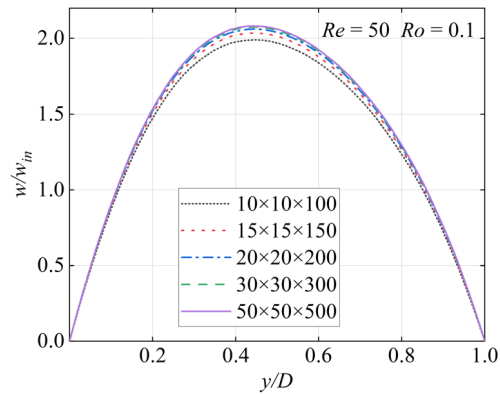

Figure 3：Grid independence verification

## 3 result

### 3.1 Main Flow Characteristics of Incompressible Flow in Rotating Channel

To investigate the flow characteristics of compressible rotating flows, we initiate our study by examining the velocity profiles of incompressible flows at various rotation number. The channel's inlet Reynolds number ($Re_{in}$) is set to 50, the inlet Mach number ($Ma_{in}$) is $2.3 \times 10^{-8}$, and the rotation radius ratio ($r/D$) is 100.

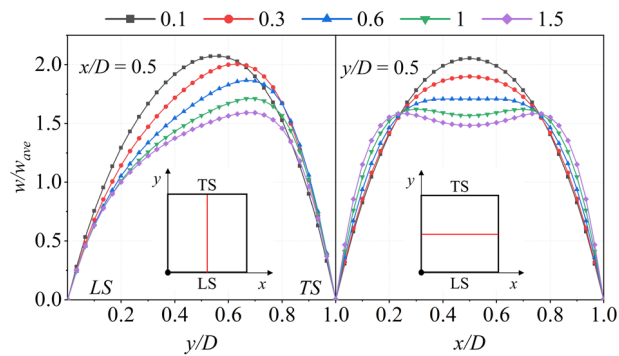

Figure 4: Comparison of Mid-Section Velocity Profiles in Incompressible Channels at Different Rotation Number

The main flow velocity profiles distributions at different rotation number are illustrated in Figure 4. With an increase in rotation number, the maximum velocity within the cross-section gradually decreases, and the fluid shifts towards both sidewalls. Furthermore, as the rotation number further escalates, the cross-sectional velocity exhibits a bimodal distribution, with the maximum velocity occurring in the areas on either side of the midline from the leading side to the trailing side.

The influence of the Coriolis force on the channel fluid is reflected in a shifting of the main flow velocity profile towards the higher-pressure side. As the rotation number increases, the slope at the leading side decreases gradually. Near the wall at the trailing side, the velocity gradient initially rises before diminishing. This behavior is primarily attributed to the reduction in mass flow rate at the $x/D = 0.5$ cross-sectional location with increasing rotation number, leading to a gradual decrease in velocity gradient near the trailing side. On both sidewalls, as the flow is continuously pushed towards the wall, an increase in the rotation number further amplifies the velocity gradient near the wall. Consequently, as the rotation number increases, the wall shear stress adjacent to the sidewalls monotonically increases.

Remarkably, in the dimensionless velocity profiles at the $y/D = 0.5$ cross-sectional location, all profiles for different rotation number pass through two fixed points. The uniformity of this phenomenon warrants validation across various Reynolds number.

### 3.2 Influence of Mach Number on Main Flow

To validate the impact of centrifugal force on fluid compression, the flow parameters for the research model were configured as detailed in Table 1.

Table 1: Flow Parameters of Channel at Different Inlet Mach Number for $Ro_{in} = 0.2$

| No. | $D$ (m) | $Re_{in}$ | $Ro_{in}$ | $Ma_{in}$ | $r/D$ |
|---|---|---|---|---|---|
| 1 | 0.001 | 50 | 0.2 | $2.3 \times 10^{-8}$ | 100 |
| 2 | 0.001 | 50 | 0.2 | $1.0 \times 10^{-1}$ | 100 |

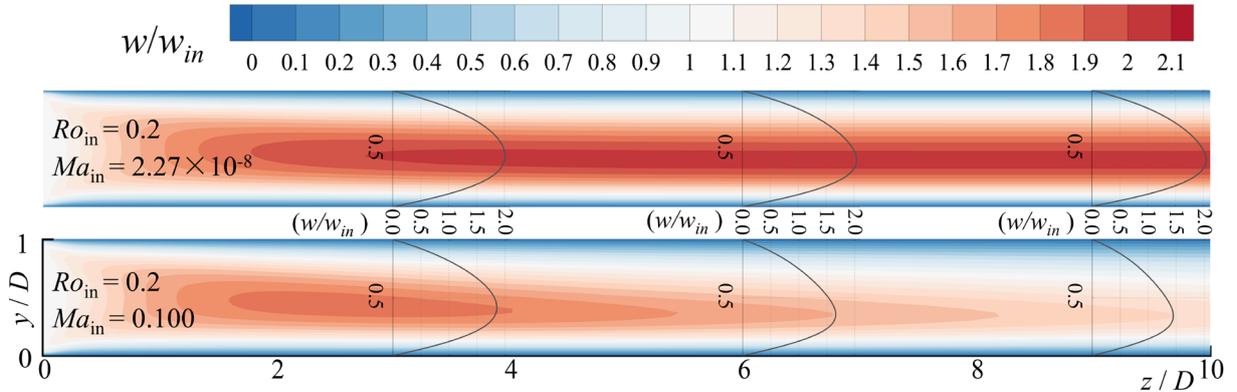

Figure 5: Comparison of Dimensionless Velocity Fields at Cross-section ($x/D = 0.5$) between Model No.1 and Model No.2

Under the same inlet Reynolds number and rotation number, Figure 5 compares the flow within channels at different inlet Mach number. In compressible flows, there is no fully developed section. As the fluid undergoes continuous compression, the velocity gradually decreases, and localized Coriolis force intensifies, leading to an increase in local rotation number. In compressible flow within rotating channels, the main velocity profile gradually shifts towards the trailing side. The primary cause of this phenomenon is the compression induced by centrifugal force, which leads to deceleration of the main flow and amplifies the relative impact of Coriolis force. Consequently, it intensifies the tendency for fluid deflection towards the trailing wall. However, further investigation is required to determine whether this deflection can be solely attributed to the deflective effect of Coriolis force or if fluid compression also contributes to the behavior of the flow.

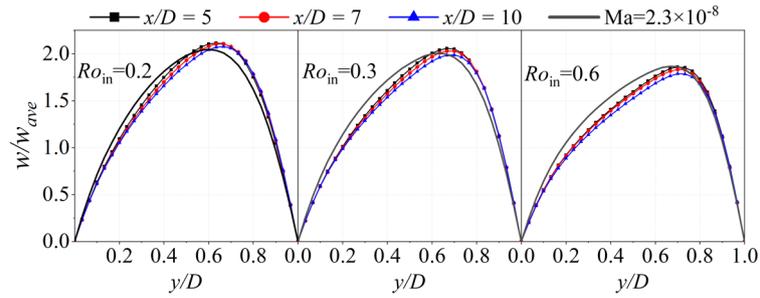

Figure 6: Variation of Main Flow Velocity Profiles along the Channel ($x/D$ = 0.5) in Compressible Channels at Different Inlet Rotation Number

Figure 6 illustrates the dimensionless main flow velocity profiles across the channel's midsection($x/D$=0.5), from the leading side to the trailing side, while maintaining consistent inlet centrifugal work number($CW$). This ensures that the centrifugal force-induced compression effect remains consistent along the channel. To mitigate the deceleration effect caused by fluid compression, the local velocities are normalized by the respective average velocities at each flow section. This normalization eliminates the impact of fluid compression on the deceleration effect.

Due to the compressive effect of rotational centrifugal forces, the main flow velocity decreases, resulting in an increase in the relative magnitude of Coriolis forces. This translates to a gradual rise in local rotating number along the channel. Upon dimensionless normalization of the local velocity field, the development of the main velocity closely follows the trend observed in incompressible channel flow with increasing rotating number. In compressible flow, as the rotating number increases along the channel, the main flow velocity profile gradually shifts towards the trailing side.

In channels with identical inlet dimensionless parameters, the compression of centrifugal forces leads to a disparity between local rotating number and the inlet conditions, resulting in alterations to the sectional flow. Thus, in compressible flow, it is less appropriate to use inlet parameters as global criteria for comparison. When comparing flows with different Mach number, opting for locally consistent dimensionless parameters would be a more reasonable choice.

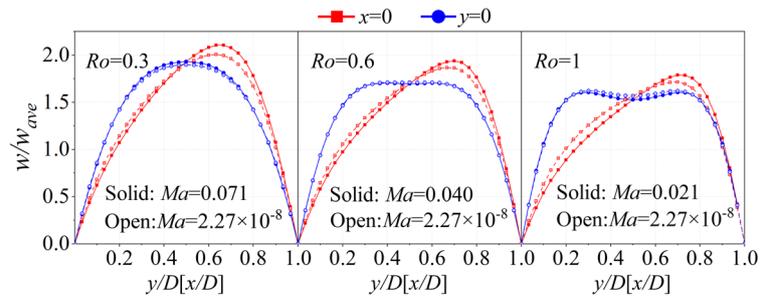

Figure 7: Comparison of Velocity Profiles at the Same Local Rotation Number Section for Different Inlet Mach Number within the Channel

Figure 7 compares the dimensionless velocity profiles across sections with the same local rotating number but varying Mach number. Observing the position of the maximum mainstream velocity, it is evident that higher Mach number yield peak surpassing that of incompressible flow, with the maximum velocity position shifting toward the trailing side. Regarding this shift toward the trailing side, it exhibits greater similarity with the flow pattern associated with higher rotating number, while the higher peak velocity align more with the pattern observed at lower rotating number. Therefore, this developmental pattern of flow seems not to be solely attributed to the influence of rotational Coriolis forces.

Furthermore, sections with higher Mach number exhibit distinct differences in their dimensionless velocity profiles compared to the incompressible state. In Figure 7, the compressible section's velocity profile near the maximum velocity is sharper, and the velocity profile near the leading side appears to be influenced by a certain compressibility effect. The velocity profile at the leading side exhibits a more linear growth, suggesting that it is influenced by a compressibility-related effect. The velocity profiles at leading side is not as full as in the mainstream velocity as observed in sections with lower Mach number.

In general, if the sole cause of longitudinal compression in channel flow is rotational centrifugal forces, variations in velocity profile along the flow can be attributed to the increasing relative magnitude of Coriolis forces. However, our current findings indicate that the enhanced effect of Coriolis forces alone cannot fully explain phenomena such as the relative increase in maximum velocity. Therefore, it can be postulated that in this straightforward rotating compressible flow, an additional force must be at play, resulting in the disparities observed in velocity profiles between compressible and incompressible flows under identical local rotational conditions. Further investigation into this matter will be conducted in Chapter 4.

### 3.3 Influence of Mach Number on Wall Shear Stress

To comprehensively investigate the variation of wall shear stress with Mach number within the channel, we conducted a study on channels with different rotational speeds across three distinct rotation number. The working medium employed in this study was an ideal gas, while maintaining a Reynolds number of 50. The inlet temperature was set to 300K, and the channel's rotational radius ratio was $r/D = 100$. We adjusted the gas pressure to modify the flow Mach number. The specific model conditions are presented in Table 2.

Table 2: Inlet Pressure and Rotational Speed for the Investigated Models at Three Rotation Number

| $Ro=0.20$—$P_{in}(Pa)$ | $Ro=0.40$—$P_{in}(Pa)$ | $Ro=0.60$—$P_{in}(Pa)$ | $n(rad/s)$ |
|---|---|---|---|
| $1\times10^{10}$ | $2\times10^{10}$ | $3\times10^{10}$ | $1.58\times10^{-3}$ |
| 10000 | 20000 | 30000 | $1.58\times10^{3}$ |
| 5000 | 10000 | 20000 | $3.00\times10^{3}$ |
| 4000 | 8000 | 16000 | $3.74\times10^{3}$ |
| 3000 | 6000 | 12000 | $4.73\times10^{3}$ |
| 2000 | 4000 | 8000 | $6.31\times10^{3}$ |
| 1500 | 3000 | 6000 | $7.36\times10^{3}$ |
| 1300 | 2600 | 3900 | $7.88\times10^{3}$ |

From the variations in the dimensionless velocity profiles(Figure 7), it becomes evident that wall shear stress is influenced by the compressibility of the fluid. The comparison presented in the figure above illustrates the dimensionless wall shear stress at the leading side ($f_{LS}$) and trailing side ($f_{TS}$) for three distinct rotation number: 0.20, 0.30, and 0.60. Additionally, the overall dimensionless wall shear stress ($f$) for the channel is included for reference.

$$f = \frac{\tau_w}{\frac{1}{2}\rho w^2} \tag{3.1}$$

Where, $\tau_w$ is the wall shear stress and $1/2\rho w^2$ is the average dynamic pressure of the section.

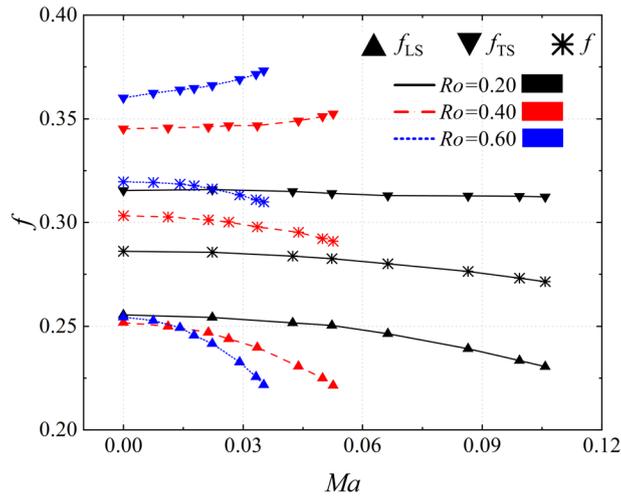

Figure 8: Variation Trend of Dimensionless Wall Shear Stress with Mach Number for Different Rotation Number

As depicted in Figure 8, under the conditions of lower Mach number in the incompressible regime, the channel's overall drag coefficient increases with an augmentation in the rotation number, aligning with the findings of Fann (1992)[11]. Within the range of $Ro$ = 0.2-0.6, minimal variations are observed in the wall shear stress near the leading side, while the wall shear stress near the trailing side escalates rapidly as the rotation number increases.

The wall shear stress on the leading side of the channel is most notably influenced by the Mach number. Due to the rotational compression effect, the wall shear stress on the leading side diminishes. Under the condition of a rotational speed of 7880 rad/s, the dimensionless wall shear stress for $Ro$ = 0.2, $Ro$ = 0.4, and $Ro$ = 0.6 are reduced by 10%, 12%, and 13%, respectively, compared to the incompressible state. This reduction is primarily attributed to the decrease in velocity gradient near the leading side, which occurs in channels with higher Mach number. Consequently, with an increase in Mach number, the dimensionless wall shear stress on the leading side decreases. At the same rotational speed, although the local Mach number decreases with higher rotation number, the decrease in wall shear stress on the leading side is particularly evident for Ro = 0.6.

Conversely, the velocity profile on the trailing side steepens as the Mach number increases, resulting in an elevation of the dimensionless wall shear stress. However, this increase is notably less pronounced than the reduction observed on the leading side. Given that the reduction in wall shear stress on the leading side is more significant compared to the variations on the trailing and lateral sides, the overall drag coefficient of the channel gradually decreases with an increasing Mach number.

### 3.4 Influence of Mach Number on Secondary Flow in Channels

Figure 9 illustrates the dimensionless velocity distribution of secondary flow within sections at three rotational number: 0.2, 0.4, and 0.6, and varying local Mach number. On the whole, the influence of increasing Mach number on secondary flow structure is not prominent; the sectional secondary flow still exhibit the characteristic double-vortex pattern. However, there is a slight impact of higher Mach number on the intensity of sectional secondary flow. The y-directional velocity at both the center and sides of the sections are enhanced with increasing Mach number.

Moreover, Mach number affect the peak location of the secondary flow. Elevated Mach number result in a more pronounced rearward shift of the secondary flow peak toward the trailing side. This phenomenon primarily arises due to the influence of Coriolis forces driving the secondary flow within the rotating sections. Hence, the intensity and peak location of the secondary flow at the section's center are closely related to the peak and position of the mainstream velocity. Considering the influence of Mach number on the mainstream, where higher Mach number lead to larger peak velocities and their shift toward the trailing side, the corresponding alterations in secondary flow peaks are expected.

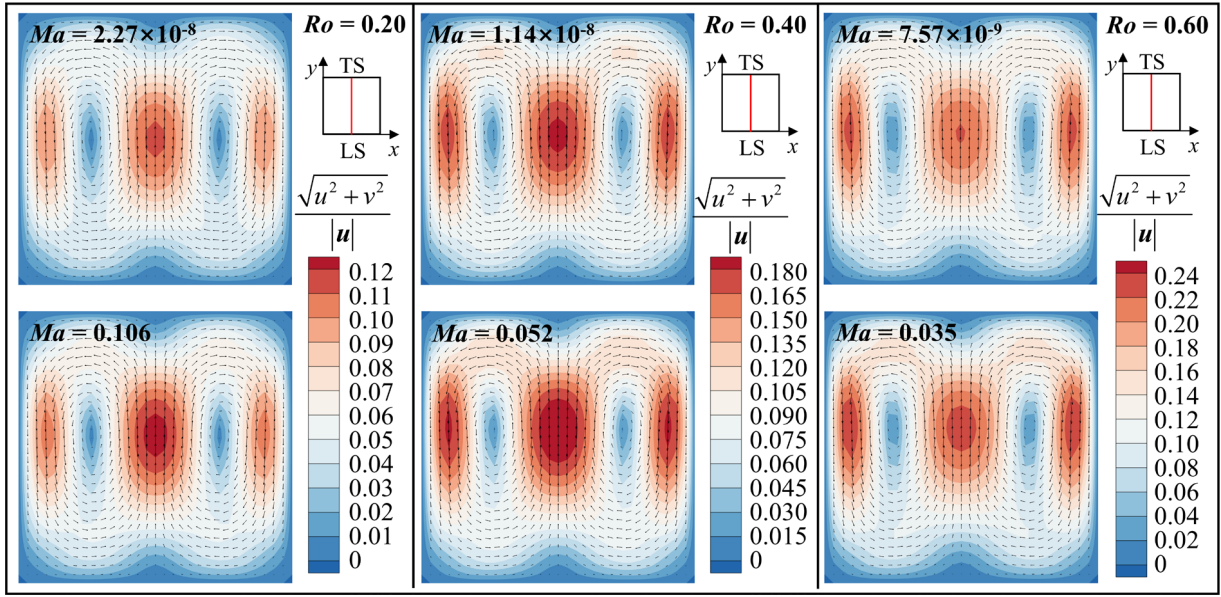

Figure 9: Dimensionless Velocity Distribution of Secondary Flow across Different Mach Number at Rotation Number 0.2, 0.4, and 0.6

## 4 Discussion

### 4.1 Effect of Mach Number-Induced Buoyancy Force

In comparison to incompressible channel, channel with higher inlet Mach number do not introduce new rotational forces. Therefore, the primary cause for the flow variations should stem from the fact that, in channel with higher Mach number, certain rotational forces' distributions exhibit significant differences from those in incompressible flow. Analyzing the key reasons behind the flow alterations from the perspective of the governing equations can provide further insights. As the Mach number studied in this work are all below 0.3, they fall within the realm of weakly compressible (subsonic) flows, without encountering strong compression effects such as shock waves. Consequently, the momentum equation can be expressed as follows:

$$\rho(\boldsymbol{u}\cdot\nabla)\boldsymbol{u} = -\nabla p + \mu\nabla^2\boldsymbol{u} - \rho[2\boldsymbol{\Omega}\times\boldsymbol{u} + \boldsymbol{\Omega}\times(\boldsymbol{\Omega}\times\boldsymbol{r})] \quad (4.1)$$

At the same rotational radius position, selecting the flow state at a point "$a$" as the reference point, the equation can be transformed into:

$$\rho(\boldsymbol{u}\cdot\nabla)\boldsymbol{u} = -\nabla p + \mu\nabla^2\boldsymbol{u} - \rho_a[2\boldsymbol{\Omega}\times\boldsymbol{u} + \boldsymbol{\Omega}\times(\boldsymbol{\Omega}\times\boldsymbol{r})] + (\rho_a - \rho)[2\boldsymbol{\Omega}\times\boldsymbol{u} + \boldsymbol{\Omega}\times(\boldsymbol{\Omega}\times\boldsymbol{r})] \quad (4.2)$$

Here, we transform the volume force term at the same rotational radius position using the fluid state at point "$a$". Consequently, an additional force term arises at the same rotational radius position due to the differences in volume force caused by variations in fluid density. Here, our primary focus is on the flow in the z-direction. In a steady state, the $z$-direction momentum equation can be represented as:

$$u\frac{\partial w}{\partial x} + v\frac{\partial w}{\partial y} + w\frac{\partial w}{\partial z} = -\frac{1}{\rho}\frac{\partial p}{\partial z} + \frac{\mu}{\rho}\left(\frac{\partial^2 u}{\partial x^2} + \frac{\partial^2 v}{\partial y^2} + \frac{\partial^2 w}{\partial z^2}\right) + \frac{\rho_a}{\rho}\left(2\Omega v + \Omega^2 r\right) - \frac{\rho_a - \rho}{\rho}\left(2\Omega v + \Omega^2 r\right) \quad (4.3)$$

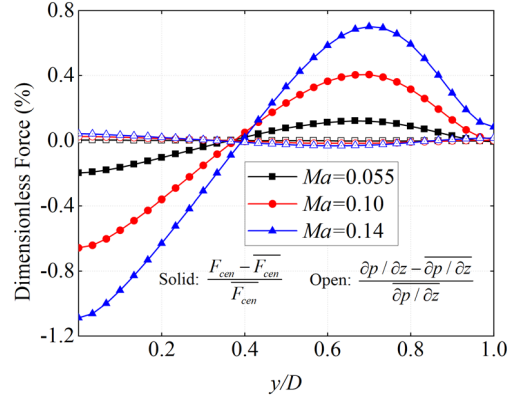

Figure 10: Variation of Dimensionless Density and Pressure Gradient at Midsection ($x/D$=0.5) across Different Inlet Mach Number in the Channel

In the expression of the equation, our primary focus lies on the influence of density variation across the cross-section of the rotating channel. However, within this equation, the effects of pressure gradient and centrifugal forces are complex. To attempt the simplification of the momentum equation, we compare the relative magnitudes of pressure gradient and centrifugal forces, as illustrated in Figure 10. In the case of Ro = 0.2, non-dimensionalization of the centrifugal force and pressure gradient terms within the channel is performed. We observe that compared to the variations in cross-sectional centrifugal force, the variations in cross-sectional pressure gradient are nearly negligible. As a result, within the rotating channel, we posit that at the same radial position, the pressure gradient along the z-direction can be considered constant across different locations. Building upon this assumption, we can further simplify the equation by representing the combination of pressure gradient and body force terms:

$$-\frac{1}{\rho}\frac{\partial p}{\partial z}+\frac{\rho_a}{\rho}\left(2\Omega v+\Omega^2 r\right)=C_z \tag{4.4}$$

Therefore, the equation in the z direction is expressed as:

$$u\frac{\partial w}{\partial x}+v\frac{\partial w}{\partial y}+w\frac{\partial w}{\partial z}=\frac{\mu}{\rho}\left(\frac{\partial^2 u}{\partial x^2}+\frac{\partial^2 v}{\partial y^2}+\frac{\partial^2 w}{\partial z^2}\right)-\frac{\rho_a-\rho}{\rho}\left(2\Omega v+\Omega^2 r\right)+C_z \tag{4.5}$$

Hence, the variation in fluid inertia can be represented using the buoyancy force term generated due to density changes and the viscous force term. In the fully developed stage of incompressible flow, where there is no density variation in the flow field, the driving force is the volumetric force term. The balance is maintained between the z-direction pressure gradient, viscous force term, and volumetric force term.

As the inlet Mach number progressively increases, density variations arise across the cross-section. In flow fields devoid of volumetric forces or where the volumetric force are minor (such as gravitational fields), the impact of these density changes at low Mach number (Ma < 0.3) is practically negligible. However, in flow fields dominated by strong centrifugal forces like those in rotating machinery, the influence of subtle local density discrepancies due to Mach number effects can give rise to significant buoyancy forces directed towards higher rotational radius. This phenomenon leads to a acceleration near the nondimensional velocity peak in compressible flow under the same rotation rate.

For simplification purposes, if we disregard deceleration in the flow direction and assume a quasi-equilibrium state at each cross-section along the flow, a scenario often encountered in practice, then to counterbalance the intensified centrifugal forces, velocity gradients further increase around the maximum velocity location. As a result, the velocity profile near the velocity peak becomes even sharper.

## 4.2 Influence of Coriolis Force on Buoyancy Force

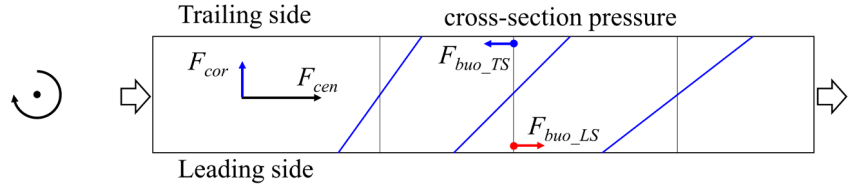

Figure 11: Schematic of Cross-Sectional Pressure Distribution in Rotating Channel

In a rotating channel, the influence of Coriolis forces also leads to pressure variations within the flow cross-section, as illustrated in Figure 11. In previous studies, the pressure differences generated by Coriolis forces at the leading and trailing sides of the flow cross-section were almost negligible and did not significantly affect the flow. However, in compressible flow, as the mainstream Mach number increases due to the influence of rotation, the pressure differences between the leading and trailing sides also increase. When the pressure difference caused by Coriolis force become significant enough, the impact of density variations within the cross-section on the flow cannot be overlooked.

In a two-dimensional rotating channel, when focusing solely on the pressure difference induced by Coriolis forces at the leading and trailing sides, leading to density variations, the pressure difference between these sides due to Coriolis force can be expressed as follows:

$$p_{TS} - p_{LS} = \int_0^d 2\rho\Omega u \, dy \tag{4.6}$$

To estimate the pressure difference between the leading and trailing sides, the integral relation in the above equation can be transformed using average parameters within the cross-section. The pressure difference between the leading and trailing sides can be approximated as follows:

$$p_{TS} - p_{LS} = 2\bar{\rho}\Omega\bar{u}d \tag{4.7}$$

In an unheated channel, the temperature variation within a cross-section perpendicular to the flow direction can be neglected. Thus, the density difference between the leading and trailing sides can be expressed as:

$$\frac{\rho_{TS} - \rho_{LS}}{\bar{\rho}} = \frac{2\Omega\bar{u}d}{R_g T} \tag{4.8}$$

Simplifying the above expression 4.8:

$$\frac{\rho_{TS} - \rho_{LS}}{\bar{\rho}} = 2k \cdot \frac{\Omega d}{u} \cdot \frac{u^2}{kR_g T} = 2k \cdot Ro \cdot Ma^2 \tag{4.9}$$

The buoyancy parameter (*Buo*), caused by the Coriolis force within the cross-section under rotation can be expressed as:

$$Buo = \frac{\Delta\rho}{\rho} \cdot Ro^2 \cdot \frac{r}{D} \tag{4.10}$$

The density variation in expression 4.10 is represented by the density variation within the cross-section caused by the Coriolis force (expression 4.9). Therefore, the buoyancy parameter generated by the Coriolis force can be expressed as expression 4.11:

$$Buo = 2k \cdot Ma^2 \cdot Ro^3 \cdot \frac{r}{D} \tag{4.11}$$

The expression 4.11 reveals that the dimensionless buoyancy force between the leading and trailing sides can be expressed as the product of cubic of the rotation number and the square of the Mach number. Thus, even when maintaining consistent local Reynolds number and rotation number, adjusting the Mach number can not only alter the compressibility effects of centrifugal force but also enhance the compressibility effects of Coriolis force. However, in cases where the sectional density remains uniform, centrifugal force would be uniformly compressed as well. When compression occurs gradually, each state can be approximated as a quasi-equilibrium state, with flow development predominantly influenced by the amplification of the Coriolis force. Nonetheless, with the increase in main flow Mach number, the compressibility effect of the Coriolis force will also escalate rapidly. The density near the trailing side is slightly higher than that near the leading side. Nevertheless, the influence of the volume force in the rotating channel is significant, for instance, at $Ro = 0.2$ and $Ma = 0.1$, $\Omega^2 r = 6.2\times10^6 m/s^2$. Therefore, even an extremely subtle density difference can have a substantial impact on the flow dynamics. Therefore, the imbalance between the pressure gradient and centrifugal force within the cross-section leads to the generation of buoyancy force.

**4.3 Verification of the Effects of Compressibility in Rotating Channel Flows**

To ascertain that the fundamental cause of flow variation is due to density changes, this study manipulated the computational model by switching the inlet and outlet positions. Fluid was made to flow from the higher radius position and exit at the lower radius position. In this scenario, the centrifugal force direction is opposite to the flow direction, implying that the effects resulting from increased density should be precisely opposite to the phenomena within the outward radial flow channel. As shown in Figure 12, a comparison was made between the mid-section velocity profiles of the outward radial flow channel and the inward radial flow channel for $Ro = 0.2$. When the channel's Mach number is low, the velocity profiles within the fully developed section of the inward and outward radial flow channels are symmetric. This confirms that in incompressible flow, the effects of the centrifugal force can be entirely equated to pressure changes, and their impact on the flow can be almost neglected. The inflow channel experiences the deviation caused by the Coriolis force direction, resulting in the main flow velocity shifting toward the leading side.

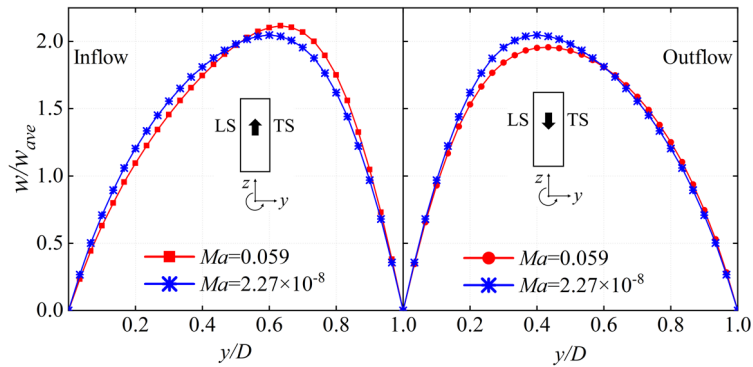

Figure 12: Comparison of Velocity Profiles between Radial Inflow and Radial Outflow Channels at Mid-Section ($x/D$=0.5) for $Ro$=0.2 and $Ma$=0.059, Compared to Incompressible Flow Profiles

As the Mach number of the inflow channel increases, as depicted in Figure 12, differences in density occur at the same radial position. Higher density is observed at the position of maximum velocity and near the leading side, resulting in a more significant additional centrifugal force component (buoyancy force) compared to the vicinity of the leading side. Due to the influence of the additional force, the velocity gradient near the maximum velocity point decreases, resulting in a smoother transition of fluid at the point of velocity extreme and a decrease in velocity peak. Near the leading side directed by the Coriolis force, the augmented effect of the centrifugal buoyancy force decelerates local fluid. If we interpret this flow state as a quasi-equilibrium state, the opposing centrifugal buoyancy force slows down the growth of the velocity gradient near the leading side and accelerates the decrease in fluid velocity near the trailing side. From the position of the velocity peak, we can also observe that, under the same local rotation number, the velocity peak of the fluid leans towards the wall with lower density(trailing side).

In the radial outward flow channel, changes in the flow velocity profile due to the influence of the Mach number confirm that sectional density alteration is the fundamental cause behind significant alterations in the flow patterns of rotating compressible pipe flow.

# 5 Conclusion

In rapidly rotating spanwise channels, volumetric forces induce a significant compressive effect along the flow direction. The substantial alterations in flow within high-speed rotating channels can be attributed to two primary factors:

1. Correlation of Centrifugal Compression with Dimensionless Parameter *CW*:

$$CW = Ro^2 \frac{r_0}{D_0} Ma^2 \tag{5.1}$$

The compressive effect generated by centrifugal forces is correlated with the dimensionless parameter *CW*. Centrifugal forces result in a decrease in fluid velocity along the streamwise direction, while the relative strength of the Coriolis force continuously intensifies. This leads to a gradual reduction in the slope of the velocity profile at the leading side and a progressive decrease in the dimensionless peak velocity of the main flow, with the peak shifting towards the trailing side in laminar flow conditions.

2. The increase in mainstream Mach number leads to compression effects near the velocity peak, and the density variations induced by the Coriolis force within the cross-sectional plane also significantly impact the flow. Specifically, the density difference caused by the Coriolis force at the leading and trailing sides can be expressed as follows:

$$\frac{\rho_{TS} - \rho_{LS}}{\rho} = 2k \cdot Ro \cdot Ma^2 \tag{5.2}$$

As the Mach number increases and the compression by the Coriolis force generates slight density disparities within the cross-section, pronounced centrifugal effects emerge under strong centrifugal forces, resulting in substantial centrifugal additional force, or buoyancy force. In the case of outward radial flow channels, where the fluid density is lower at certain locations within the cross-section, the buoyancy force acts toward the center of rotation. The buoyancy parameter generated by the Coriolis force can be expressed as:

$$Buo = 2k \cdot Ma^2 \cdot Ro^3 \cdot \frac{r}{D} \tag{5.3}$$

Compared to incompressible rotating channels, this effect diminishes the dimensionless velocity of local fluid. Conversely, at positions with higher cross-sectional density, the dimensionless velocity increases. In the case of inward radial flow channels, the situation is reversed.

The wall shear stress is also influenced by the Mach number. In the outward radial flow channels, the dimensionless velocity gradient near the leading side decreases with an increase in Mach number, leading to a reduction in the wall shear stress at the leading as the compression effects intensify. Conversely, the pattern of change in wall shear stress at the trailing side is opposite. Overall, an increase in Mach number leads to a decrease in the overall channel flow resistance coefficient. Rotation number is another important parameter affecting wall shear stress variation, where higher rotation number magnify the impact of Mach number on wall shear stress.

The cross-sectional secondary flow structure within the rotating channel remains relatively unaffected by the Mach number within the studied range of Reynolds and rotation number. The typical double-vortex pattern is retained. However, influenced by the location and intensity of the mainstream peak, the secondary flow strength increases, and the peak position shifts towards the trailing side.

Within the scope of this study, both the compressive effects of centrifugal force and the density variations caused by velocity changes and Coriolis compression are influenced by the Mach number. The increase in Mach number is the fundamental reason behind the significant differentiation between fluid flow within rotating channels and incompressible channels.

In contrast to static conditions, the conventional conclusion that fluid compressibility effects can be neglected for Ma < 0.3 is not applicable under rotational states. Defining a critical Mach number distinguishing compressible from incompressible flow becomes challenging under rotation, as the compressive effects involve not only the centrifugal force's impact but also the buoyancy force resulting from the combined influence of Mach number compressibility and Coriolis compression within the flow section. Therefore, various rotation number(Ro), radius ratios($r/D$), and Mach number($Ma$) collectively influence the channel's compressive behavior. Investigating the influence of Mach number on flow within rapidly rotating channel across different rotational conditions holds substantial significance.

## Declaration of Interests

The authors report no conflict of interest.

## Author Contributions

**Junxin Che**: Data curation (equal); Formal analysis (equal); Resources (equal);Writing - original draft (equal);Writing - review & editing (equal). **Ruquan You**: Formal analysis (equal); Funding acquisition (equal); Project administration (equal); Supervision (equal). **Wenbin Chen**: Conceptualization (equal); Investigation (equal);Methodology (equal). **Haiwang Li**: Supervision (equal); Writing – review & editing (equal).

## Acknowledgments

This work was sponsored by Beijing Nova Program [No. 20220484129]；

This work was supported by Beijing Municipal Natural Science Foundation [No. 3222034];

This work was supported by National Science Fund for Distinguished Young Scholars [No. 52225602];

This work was supported by "the Fundamental Research Funds for the Central Universities".

## Data Availability

The data that support the findings of this study are available within the article.

## References

1. J. P. Johnston, R. M. Halleent, andD. K. Lezius, "Effects of spanwise rotation on the structure of two-dimensional fully developed turbulent channel flow," Journal of Fluid Mechanics **56**, 533 (1972).

2. R. Kristoffersen, andH. I. Andersson, "Direct simulations of low-Reynolds-number turbulent flow in a rotating channel," Journal of Fluid Mechanics **256**, 163 (1993).

3. P. H. Alfredsson, andH. Persson, "Instabilities in channel flow with system rotation," Journal of Fluid Mechanics **202**, 543 (1989).

4. D. K. Lezius, andJ. P. Johnston, "Roll-cell instabilities in rotating laminar and trubulent channel flows," Journal of Fluid Mechanics **77**, 153 (1976).

5. C. G. Speziale, "Numerical study of viscous flow in rotating rectangular ducts," Journal of Fluid Mechanics **122**, 251 (1982).

6. C. G. Speziale, andS. Thangam, "Numerical study of secondary flows and roll-cell instabilities in rotating channel flow," Journal of Fluid Mechanics **130**, 377 (1983).

7. Y. Mori, T. Fukada, andW. Nakayama, "Convective heat transfer in a rotating radial circular pipe (2nd report)," International Journal of Heat and Mass Transfer **14**, 1807 (1971).

8. W. Morris, andT. Ayhan, "Observations on the influence of rotation on heat transfer in the coolant channels of gas turbine rotor blades," Proceedings of the Institution of Mechanical Engineers **193**, 303 (1979).

9. R. Siegel, "Analysis of Buoyancy Effect on Fully Developed Laminar Heat Transfer in a Rotating Tube," Journal of Heat Transfer **107**, 338 (1985).

10. C. Soong, andG. Hwang, "Laminar mixed convection in a radially rotating semiporous channel," International journal of heat and mass transfer **33**, 1805 (1990).

11. S. Fann, andW.-J. Yang, "Hydrodynamically and thermally developing laminar flow through rotating channels having isothermal walls," Numerical Heat Transfer **22**, 257 (1992).

12. W.-M. Yan, "Effects of wall transpiration on mixed convection in a radial outward flow inside rotating ducts," International Journal of Heat and Mass Transfer **38**, 2333 (1995).

13. J.-C. Han, andY. M. Zhang, "Effect of Uneven Wall Temperature on Local Heat Transfer in a Rotating Square Channel With Smooth Walls and Radial Outward Flow," Journal of Heat Transfer **114**, 850 (1992).

14. J.-J. Hwang, andD.-Y. Lai, "Three-Dimensional Laminar Flow in a Rotating Multiple-Pass Square Channel With Sharp 180-Deg Turns," Journal of Fluids Engineering **120**, 488 (1998).

15. K. Nakabayashi, andO. Kitoh, "Turbulence characteristics of two-dimensional channel flow with system rotation," Journal of Fluid Mechanics **528**, 355 (2005).
16. G. Brethouwer, "Linear instabilities and recurring bursts of turbulence in rotating channel flow simulations," PHYSICAL REVIEW FLUIDS **1**, (2016).
17. G. Brethouwer, "Statistics and structure of spanwise rotating turbulent channel flow at moderate Reynolds numbers," Journal of Fluid Mechanics **828**, 424 (2017).
18. O. Grundestam, S. Wallin, andA. V. Johansson, "Direct numerical simulations of rotating turbulent channel flow," Journal of Fluid Mechanics **598**, 177 (2008).
19. M. I. Khan, "Nonlinear stability characteristics of rotating plane channel Poiseuille flow with the outcome of nonlinear evolution equation," Aip Advances **11**, (2021).
20. Z. Xia, Y. Shi, andS. Chen, "Direct numerical simulation of turbulent channel flow with spanwise rotation," Journal of Fluid Mechanics **788**, 42 (2016).
21. Z. Xia, Y. Shi, Q. Cai, M. Wan, andS. Chen, "Multiple states in turbulent plane Couette flow with spanwise rotation," Journal of Fluid Mechanics **837**, 477 (2018).
22. X. I. A. Yang, andZ. Xia, "Bifurcation and multiple states in plane Couette flow with spanwise rotation," Journal of Fluid Mechanics **913**, A49 (2021).
23. S. Zhang, Z. Xia, andS. Chen, "Flow structures in spanwise rotating plane Poiseuille flow based on thermal analogy," JOURNAL OF FLUID MECHANICS **933**, (2022).
24. T.-M. Liou, M.-Y. Chen, andK.-H. Chang, "Spectrum analysis of fluid flow in a rotating two-pass duct with detached 90° ribs," Experimental Thermal and Fluid Science **27**, 313 (2003).
25. A. Di Sante, andR. A. Van den Braembussche, "Experimental study of the effects of spanwise rotation on the flow in a low aspect ratio diffuser for turbomachinery applications," EXPERIMENTS IN FLUIDS **49**, 585 (2010).
26. J. Visscher, andH. I. Andersson, "Particle image velocimetry measurements of massively separated turbulent flows with rotation," Physics of Fluids **23**, 075108 (2011).
27. F. Coletti, D. L. Jacono, I. Cresci, andT. Arts, "Turbulent flow in rib-roughened channel under the effect of Coriolis and rotational buoyancy forces," Physics of fluids **26**, (2014).
28. R. You, H. Li, Z. Tao, andK. Wei, "Measurement of the Mean Flow Field in a Smooth Rotating Channel With Coriolis and Buoyancy Effects," Journal of Turbomachinery **140**, (2018).
29. J. Che, R. You, F. Zeng, H. Li, W. Chen, andZ. Tao, "Study of the Effect of a Novel Dimensionless Parameter--the Centrifugal Work Number (CW), on Spanwise Rotating channel Low-speed Compressible Flow," arXiv preprint arXiv:2305.17908 (2023).